\documentstyle[prl,aps,epsfig,floats]{revtex}

\newcommand\beq{\begin{equation}}
\newcommand\eeq{\end{equation}}
\newcommand\bea{\begin{eqnarray}}
\newcommand\eea{\end{eqnarray}}

\newcommand\nonum{\nonumber}
\newcommand\sqpi{\sqrt{\pi}}
\newcommand\tphi{{\tilde\phi}}
\newcommand\on{\omega_n}
\newcommand\bi{\begin{itemize}}
\newcommand\ei{\end{itemize}}

\begin{document}

\draft

\textheight=23.8cm
\twocolumn[\hsize\textwidth\columnwidth\hsize\csname@twocolumnfalse\endcsname

\title{\Large Effective action and interaction energy of
coupled quantum dots }
\author{\bf Sourin Das and  Sumathi Rao } 
\address{\it  
Harish-Chandra Research Institute,
Chhatnag Road, Jhusi, Allahabad 211019, India}

\date{\today}
\maketitle

\begin{abstract}

We obtain the effective action of tunnel-coupled quantum dots, 
by modeling the system as a Luttinger liquid with multiple barriers.
For a double dot system, we find that the resonance conditions for 
perfect conductance form a hexagon in the plane of the two gate
voltages controlling the density of electrons in 
each dot. We also explicitly obtain the functional dependence of the
interaction energy and peak-splitting
on the gate voltage  controlling tunneling between the dots
and their  charging energies. 
Our results are in good
agreement with recent experimental results, from which we obtain
the Luttinger interaction parameter $K=0.74$.

\end{abstract}
\vskip .5 true cm

\vskip.5pc
]
\vskip .5 true cm

Coulomb blockade {CB}\cite{GRABERT}
in quantum dots  
has gained in importance in recent years, due to its
potential for application as single electron transistors.
Recent experiments\cite{DD1,DD2,DD3}
on coupled quantum dots have revealed   
a variety of features,
the most salient of which is
the conductance peak splitting controlled by inter-dot
tunneling. The peaks split into two for a double dot
system and into three for a triple dot system.  

Earlier theoretical studies\cite{CD} on coupled dots
studied the effect 
of tunneling and inter-dot capacitances 
on the CB, but only a 
few\cite{MGB,GH,LJ} focused on junctions with only one
or two channels. Motivated by a) the double dot  experiments,
b) the analysis of Matveev\cite{MATVEEV}, which maps a
quantum dot formed by a narrow constriction that allows only
one transverse channel to enter the dot, to a one-dimensional
wire model and  
c) recent numerical evidence\cite{ASSA} that justifies modeling
of quantum dots by Luttinger liquids, we
model the coupled dot system as
a one-dimensional Luttinger liquid (LL) with large barriers and  
study the effective action in the coherent limit.

We obtain a simple expression
for the interaction energy in terms of the charging energies of the
two dots and the gate voltage controlling the inter-dot
tunneling, and find that tunneling between the dots splits the 
conductance maxima.
With (without) inter-dot tunneling, the conductance (resonance) maxima,
for a double dot system, 
form a rectangular (hexagonal) grid in the plane of the two gate voltages
controlling the density of the electrons. 
The peak splitting 
grows to a maximum when the two dots merge into a single dot.
From the experiments, we estimate the interaction energy
and  also the values for the Luttinger parameters. 
 
Although a classical analysis with an inter-dot capacitance
reproduces the experimental results qualitatively, the value of the required 
inter-dot capacitance is unrealistically large\cite{DD1}.
Golden and Halperin\cite{GH} obtain their results by mapping the 
two dot model to a single dot model. 
Here, we directly model the experimental set-up as a one-dimensional
system and obtain the effective action
exactly, by integrating out Gaussian degrees of freedom.
Hence our results give the interaction energy directly in
terms of the microscopic variables and applied voltages. We find that in
the weak tunneling limit, the peak
splitting is proportional to the square-root of the tunneling
conductance and in the strong tunneling limit, the deviation of
the splitting factor from unity is proportional to the square-root
of the deviation of the conductance (measured in units of $2e^2/h$)
from unity.

\noindent{\it Model and effective action for coupled  quantum dots}

Our model for a dot is essentially a short length of LL
wire with barriers at either end, controlling the tunneling to and
from the dot. 
The two dot system is  shown in Fig.(1), 
with three 
barriers in the LL wire between the two leads.
The barriers are chosen to be  $\delta$-functions with strengths $G_i$
and $J$. (Extended barriers do not change the results\cite{GH}.)
The electrons in the dot interact via a short range (Coulomb-like)
interaction described by the Hamiltonian $H=-v_F \int dx 
[\psi_L^{\dagger}i\partial_x \psi_L - (L\leftrightarrow R)] +g \int dx 
\rho(x)^2$ where $\rho(x) = \psi^{\dagger}\psi$
is the electron density and $\psi = \psi_L e^{-ik_F x} +\psi_R e^{ik_Fx}$.
Here, $\psi_R$ and $\psi_L$ stands for fermion fields linearised
about the left and right Fermi points.
The electrons in the leads are free.

The model is bosonised via the standard transformation
$\psi_\nu = \eta_\nu e^{2i\sqpi \phi_\nu}$ where $\nu=R,L$ and
$\phi_\nu$ are the bosonic fields. $\eta_\nu$ are the Klein factors 
that ensure anti-commutation of the fermion fields. 
The Lagrangian density for the  bosonic field in 1+1 dimensions
is ${\cal L}(\phi;K,v) =
(1/2Kv)(\partial_t \phi)^2 -(v/2K)(\partial_x\phi(x))^2$,
where $K\sim(1+g/\pi v_F)^{-1/2}$ and $v\sim v_F(1+g/\pi v_F)^{1/2}$.
The bosonised action is now given by\cite{KF,LAL,RS}
\beq
S = \int d\tau [S_{leads} + S_{dots} + S_{gates}] ~, 
\eeq
with 
$S_{leads} = (\int_{-\infty}^{-d_1} + \int_{d_2}^\infty ) 
dx {\cal L}(\phi;K_L=1,v_F) $, 
$S_{dots} =  \int_{-d_1}^{0}  
dx {\cal L}(\phi;K_1,v_1)
+\int_{0}^{d_2} dx {\cal L}(\phi;K_2,v_2) 
+V+ J\cos(2\sqpi\chi)$ and  
$S_{gates} = (g_1/\sqpi)  \int_{-d_1}^{0} dx (\chi-\phi_1) 
  +  (g_2/ \sqpi) \int_{0}^{d_2} dx (\phi_2-\chi)~$,
where $V=\Sigma_{i=1}^2 
G_i\cos (2\sqpi\phi_i + (-1)^i 2k_F d_i)$.
Here 
$G_i$ and $J$ are  the junction barriers defining
the dots and $\phi_i \equiv \phi((-1)^id_i)$ and
$\chi\equiv \phi(0)$ denote the boson fields at those points. 
$J$ controls
tunneling between the dots and ranges from fully open
($J=0$, single dot ) to fully closed ($J=\infty$, two decoupled dots). 
The gate voltages $g_i$ control the density of electrons
in each dot.
  
\begin{figure}[htb]
\begin{center}
\epsfig{figure=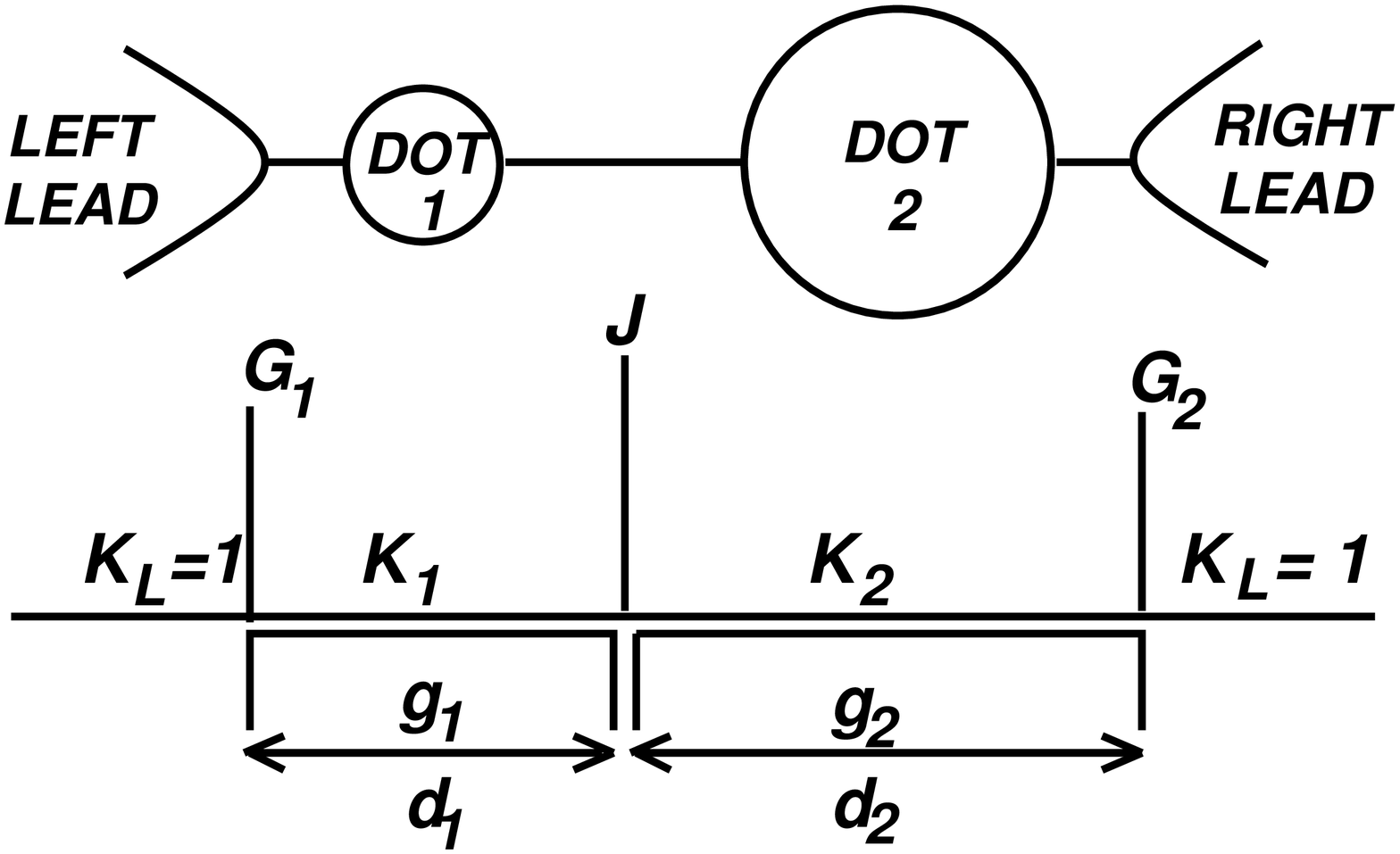,width=4.0cm}
\end{center}
\caption{Double dot system above shown schematically below
with the dots modeled as LL with $K_{i =1,2}$, 
between barriers $G_i$ and $J$, and with two gate voltages $g_i$.}
\end{figure}

In the low $T$ limit  
($T \ll \hbar v_1 /k_B d_1, \hbar v_2/k_B d_2$) 
all three barriers are  seen coherently, and
we expect the CB of the two dots
to be coupled. 
The effective action  of the system can be obtained by integrating
out all degrees of freedom except those at the positions of the
three  junction barriers, following Ref. \cite{KF,LAL} and we find that
\bea
S_{ef} &=& S_0
+ \int d\tau \{\Sigma_{i=1}^2\Large[{U_i \over 2} ( \chi - \phi_i )^2 +
(-1)^i {g_i \over \sqrt \pi} (\chi - \phi_i)\Large]  \nonum \\
&&+ V+ J \cos {(2 \sqrt \pi \chi)}\}   \label{Seff}
\eea
where 
$S_0 = \sum_{\on} |\on| (\tphi_1^2+\tphi_2^2)/2(K_L=1)$~.
The Fourier transformed tilde fields are defined by 
$
\phi_i(\tau) = \sum_{\on} e^{-i\on \tau} \tphi_{i} (\on)
$.
The $U_i = \hbar v_i/K_i d_i$ terms are the `mass' terms that suppress charge
fluctuations
on the dots and are responsible for the CB
through the dots.

\noindent {\it Weak tunneling limit}

Here the barrier
$J$ is very large, and the field $\chi$ 
is pinned at its minima -
\beq 
\chi =  n\sqpi \equiv \chi_{cl} \quad (n={\rm integer}).
\eeq 
A semi-classical expansion about these minima
using $\chi = \chi_{cl} +\chi_{q}$
to second order in the fluctuations 
yields
\bea
S_{ef} &\simeq& S_0 +  \int d\tau \{[-g_1\phi_1 + g_2 \phi_2]/\sqpi ~+~ 
\chi_{q} (g_1-g_2) / \sqrt \pi  
 \nonumber \\
&&+ \Sigma_{i=1}^2{U_i } \big[ \chi_{q}^2 + 
2\chi_{q}(\chi_{cl}-\phi_i) + (\chi_{cl} - \phi_i)^2 \big]/2
 \nonumber \\
&& +V
- (2\pi J ) \chi_{q}^2            \}     
\eea
where we have dropped all field-independent constants.
We can now integrate out the quantum fluctuations in $\chi$, since 
we have only kept quadratic terms, 
and using $J \gg U_i, g_i$, we get
\bea
S_{ef} & \simeq & S_0 + \int d\tau[\Sigma_{i=1}^2
{U_i\over 2}(\chi_{cl} -\phi_i)^2 
  + {U_1 U_2 \over 4\pi J} \phi_1 \phi_2 \nonum \\ &&
  + {(-g_1 \phi_1 +g_2 \phi_2)/ \sqpi} +
   V ]~.
\eea
Now, we define the following variables -
$n_i = (-1)^{i+1}(\chi_{cl}-\phi_i)/ \sqpi +(k_F d_i)/ \pi$
and $\theta_i = (\chi_{cl} + \phi_i)/2 + (-1)^i (k_F d_i)/2$ 
which can be interpreted as the charges and currents on the dots 1 and
2 respectively and in terms of which the action can be
rewritten simply as
\bea
S_{ef} &\simeq S_0& + \int d\tau[\Sigma_{i=1}^2
{{\cal U}_i \over 2}(n_i -n_{0i})^2 -  
E_{12}n_1 n_2 +V]  \\
{\rm where}&& \nonum \\ 
E_{12}  &=& {U_1 U_2/4J}, ~~\&~~
n_{0i} = (k_F d_i)/ \pi - g_i/{\cal U}_i~.
\label{nio}
\eea
${\cal U}_i\equiv  \pi U_i$ plays the role of the charging energy 
or CB energy because in the absence of any mixing between $n_1$ and $n_2$,
by tuning $g_i$ (equivalently $n_{0i}$), 
the dot states with $n_i$ and $n_i+1$ electrons
can be made degenerate.
{\it This is the lifting of the CB for each individual dot.}
But the 
crucial term above is the term mixing $n_1$ and $n_2$.
This tells us how the CB through one dot
is affected by the charge on the other dot.
The barrier terms can be more conveniently rewritten in terms of
the total charge of the double dot and the current through the
double dot system, when $G_1 = G_2 = G$ by defining the total
charge and current fields as
$
N = n_1 + n_2  = (\phi_2-\phi_1)/ \sqpi + ({k_F L})/\pi
$
and $\theta = (\phi_2+\phi_1)/2 + 
({k_F l})/2\sqpi$, 
where $L = d_1+d_2$ and $l=d_2-d_1$. 
With these redefinitions, the barrier term reduces to
\beq
V = 2G\cos (2\sqpi \theta) \cos \pi N~,
\eeq 
and $S_0 \rightarrow S'_0 = 
\sum_{\omega_n} |\omega_n| ({\tilde\theta}^2 
+ (\pi/4){\tilde N}^2)$.
The action is now in a form where its symmetries are manifest.

When the two dots are decoupled from each other ($E_{12}=0$
or equivalently $J \rightarrow \infty$), 
the action is symmetric under
the transformation
$
\theta \rightarrow \theta + \sqpi, ~~N \rightarrow 2N_0 - N
$. This is because $g_1$ and $g_2$ can be tuned so as to 
make $n_{01},n_{02}$ to be half-integers $q_1+1/2,q_2+1/2$ and hence,
$N_0 = n_{01} + n_{02}$ to be an integer.
Thus the two CB's are lifted and   
the four charge states, $(n_1,n_2) = (q_1,q_2), 
(q_1+1,q_2),(q_1,q_2+1)
(q_1+1,q_2+1)$ become degenerate and transport through the double dot
system is unhindered. 
We can plot the points of maximum conductance through the double-dot
system in the plane of the two gate voltages and as shown by  the
dotted lines in Fig.(2),
we get a square grid. (The figure is for identical dots; in general,
the grid is rectangular.)

\begin{figure}[htb]
\begin{center}
\epsfig{figure=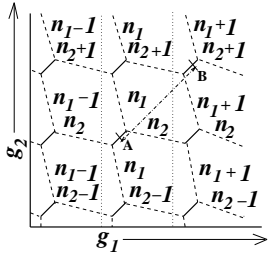,width=3.5cm}
\end{center}
\caption{Positions of the conductance maxima as a function
of the gate voltages. The solid lines denote the splitting and
the dot-dash line AB denotes the distance between two sets of split peaks.
The splitting factor $S$ is defined as twice the  ratio of the former to the
latter length.}
\end{figure} 

Now, we study the symmetries for a large finite $J$. In this case, the
CB through each dot is influenced by interaction
with the electrons in the other dot. Hence, we now find that the
CB through the double dot system  is  lifted  
for two possible sets of gate voltages - \\
$\bullet$
the states [$(q_1,q_2)$, $(q_1+1,q_2)$] and the states
[$(q_1,q_2)$, $(q_1,q_2+1)$] are degenerate simultaneously
respectively when
\beq 
n_{01} = q_1+{1\over 2} + {U_2\over 8J} q_2,      
~~~ n_{02} = q_2+{1\over 2}  + {U_1\over 8J}q_1  
\label{res1}
\eeq
This is one point where the CB is  lifted
and the conductance is a maximum. \\
$\bullet$
Similarly, the states  [$(q_1,q_2+1)$, $(q_1+1,q_2+1)$]
and [$(q_1+1,q_2)$, $(q_1+1,q_2+1)$] are degenerate simultaneously
when 
\beq
n_{01} = q_1+{1\over 2} + {U_2\over 8J} (q_2+1),      
~~~n_{02} = q_2+{1\over 2}  + {U_1\over 8J} (q_1+1)
\label{res2}
\eeq
This is another  point where  the CB  is lifted
and the conductance is a maximum. \\
Note that the original resonance which occured when all four states were
degenerate has now broken into two separate resonances, at each of
which only three states are degenerate.
Plotting this in the plane of the gate voltages, we obtain a hexagon
as shown by the dashed and solid lines in Fig.(2) in agreement with the
experiments of Blick {\it et al} in \cite{DD2}. (The figure shown
is for identical dots; otherwise, the hexagons are not perfectly
symmetric.) 
Thus, as a consequence of tunneling between the
quantum dots, the conductance maxima splits into two.

The splitting $\delta g$ of the maxima as a function
of the tunneling barrier $J$ (the solid line in Fig.(2)) can
be computed from Eqs(\ref{res1}) and (\ref{res2}), and gives  
\beq
\delta g_1 = \delta g_2
= \pi U_1 U_2/8J~ \Rightarrow \delta g ={\sqrt 2} \pi U_1 U_2/8J~.
\label{sp1}
\eeq 
Hence, the splitting  in $g_1$ and $g_2 $ is identical 
and is inversely proportional to the gate voltage
controlling the tunneling
barrier and directly proportional to the charging energies of the
two dots. 
However, when 
$J$ becomes comparable to the other energy scales in the
problem like $U_i$, the strong barrier analysis breaks down.

\noindent{\it Strong tunneling or weak barrier limit}

Here we assume that the barrier between the two dots is very
small - $i.e.$, $G_i \gg U_i \gg J$. So essentially, there is only 
one CB. However, we shall see that the spacing
of the CB terms are affected by a non-zero $J$.
We integrate
out $\chi$ (in Eq.\ref{Seff}) which is quadratic (when $J$ is ignored)
to obtain
\bea
S_{ef} &=& \Sigma_{\on}|\on| ({\tilde\theta}^2 
+ {\pi\over 4}{\tilde N}^2) + U_{ef} (N-N_0)^2/2 \nonum \\&+&
2G\cos (2\sqpi \theta) \cos \pi N
+J \cos [2\sqpi \theta + 
X\pi N + C]~ \nonum \\
&=& S_0 +V_{ef} ~.
\eea
Here, we have defined $U_{ef} =  \pi U_1U_2/(U_1+U_2)= \pi U/2$,
$X = (U_1-U_2)/(U_1+U_2)=0$,
$N_0 = k_F L/\pi - (g_1/{\cal  U}_1 +g_2/{\cal U}_2)=
k_F L/\pi -g/U_{ef}$, $C =2(g_2-g_1)/(U_1+U_2) +
(U_1+U_2)k_F l +(U_1-U_2)k_F L=0$. 
$N$ and $\theta$ have been defined earlier. $C$ and $N_0$
depend on the gate voltages and are the tunable parameters.
The second equalities are for identical dots when
$U_1=U_2=U$ and $d_1=d_2$. $g_1=g_2=g$ can be tuned by $C$.
We shall use these values below for simplicity.

Since, we continue to be in the strong barrier limit for the
$G$ term, the action has deep minima for integer $N$
(with appropriate $\theta$). But the degeneracy between these
minima is broken by the $U_{ef}$ and $J$ terms leading to just
one of them being preferred (as is the case for the usual
CB without $J$). Just as earlier, the
CB is lifted
when  
\beq
V_{ef}(\theta, N) =V_{ef}(\theta+\sqpi/2, N+1)~.
\eeq  
When $J=0$, this happens when $N_0=$ half-integer.
However, when $J\ne 0$, the above 
equality holds when 
\beq
N_0 = {1/2}  - {J/U_{ef}},{3/2}  + {J/U_{ef}},
{5/ 2}  - {J/U_{ef}} \dots,
\label{aper}
\eeq
which implies that the splitting $\delta g$  is given by
\beq
\delta g = U_{ef}-2J~.
\label{sp2}
\eeq
It is easy to see from Eq.(\ref{aper}) that
periodicity is restored when two electrons are
added to the system, but periodicity when a single
electron is added to the system is broken by the $J$ term.
This tells us that the main effect of a small barrier
within a single dot, is to change the spacing of the peaks with 
two peaks coming closer to each other and pairs of peaks
receding from each other. But the distance between pairs of  peaks
remains $2U$.
As $J$ increases towards $\infty$, the peaks which come closer
to one another slowly merge into one, and we are left with half
as many peaks, which is what one would expect when the dot
size is halved.  

\noindent{\it Discussions and conclusions}

Thus, both from the weak and strong tunneling limits,
we get a consistent picture. The splitting is zero for $J=\infty$ and
maximum when the tunneling
goes to one ($J=0$) and is just the spacing of a single dot
of length $d_1+d_2$. As a function of the tunneling barrier, the
splitting saturates  linearly (proportional to $J$)
for low barriers (strong tunneling) and 
goes as $1/J$ for weak tunneling (strong barriers).
(Our model does not include inter-dot capacitance since it is
expected to be very small for the experiment\cite{DD1}.)
In terms of the barrier conductance $G_b$, simple quantum mechanics
shows that $G_b$ (in units of $2e^2/h$) falls off from unity
as $J^2$ in the low barrier
limit and increases from zero as $1/J^2$ in the weak tunneling
limit. So in the weak tunneling limit, using Eq.(\ref{sp1}) and Fig. 2,
the splitting factor $S = \delta g/(U+\delta g) \sim 
\delta g/ U$ (for $\delta g \ll U$) $\propto 1/J \propto \sqrt G_b$.  
In the strong tunneling limit,
$S = 2\delta g/2U_{ef} = 1-2J/U_{ef}$.
So $1-S \propto J \propto \sqrt{1-G_b}$ 
Thus, in both limits, we
relate the splitting factor to the barrier conductance.
{\it This is the central result
of this letter.}  

In Fig.3, we compare our theoretical prediction of $S$ with the
experimental curve. In the weak tunneling limit, a one-parameter $\chi^2$ fit 
of our prediction to the lowest five points
(upto  $S \sim 0.3$) 
gives a goodness of fit of 84\% . In contrast, 
a linear fit to the same data ( predicted in Ref.\cite{GH}),
gives a goodness of
fit of only 75\%. Similarly, in the strong tunneling limit ($S$
above $0.7$),  
a one parameter fit of our prediction (which, in fact,
is in agreement with the scaling analysis of Ref.\cite{FLENSBERG})
gives a goodness
of fit of  96\%, again, considerably better than the 81\% for 
the logarithmic prediction in Ref.\cite{GH}. However, although our modelling 
is more realistic than earlier ones and the agreement of
our predictions with the experimental data, both  in the strong
and weak tunneling limits is impressive, better data is required
for a conclusive proof. 
Further, we predict that for the weak tunneling case, the interdot
interaction energy is directly proportional to the charging energies of the
dots and hence inversely proportional to the
sizes or capacitances of each of the individual dots.

The charging energies and the Fermi velocity obtained
from  Ref.\cite{DD1} are  $U_i \sim 400\mu eV$. and
$(v_F)_{1d}\sim 2.3 \times 10^{7} cm$.
Thus using the relation, $U=\hbar v /Kd$,
for a single dot, we find 
the LL parameters, $v_1=v_2
\sim 3.1\times 10^7 cm/sec$ and $K_1=K_2\sim.74$
for identical dots. In the weak barrier limit, where the splitting
is small, we find that $E_{12}$ (in Eq.\ref{nio}) ranges 
from $\sim 4- 40 \mu eV$, which is roughly $(10^{-2} - 10^{-1}) U$. 
The LL parameter  values remain almost
unchanged in the range of gate voltages used in the experiments.
The values  can be confirmed by studying the conductance through
the double dot system in the `high' temperature limit, 
$T\gg T_d=\hbar v/k_B d 
\sim 0.6 K$, 
when the electron
transport through each of the dots is incoherent.
The  conductance should scale as $T^{2(1/K-1)}\sim T^{0.69}$ in the 
weak tunneling case and
as  $T^{2(K-1)}\sim T^{-0.51}$ in the weak barrier case.
LL behaviour can also be probed if the conductances
are studied as a function of the size of the dots at `low'
temperatures $T\ll T_d$  since LL
theory predicts explicit size dependent power laws at low temperatures.

\begin{figure}[htb]
\begin{center}
\epsfig{figure=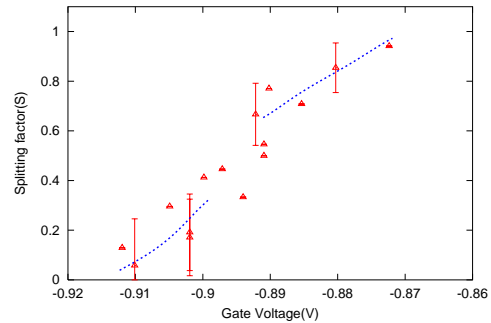,width=6.5cm}
\end{center}
\caption{Splitting factor $S$ as function of the gate voltage
controlling tunneling. The data points are from Ref.(2).
The solid line denotes
the  square-root of the measured conductance (multiplied 
by scaling factor 0.56 for best fit), and the dotted line
just denotes the measured conductance
(multiplied by 0.60 for best fit).}
\end{figure}

In conclusion, in this letter, we have obtained the effective action
of a system of coupled quantum dots, in both the weak and strong
tunneling limits. We have shown that in the presence of inter-dot tunneling,
the peaks denoting the conductance maxima split. The split is
maximum when the tunneling between the dots is maximum - i.e.,
when there is no barrier at all between the dots.
We have also computed the splitting as a function of the gate voltage
controlling the tunneling between the dots in both the
weak and strong tunneling limits, and find that the splitting factor
is proportional
to the square-root of the conductance in the weak tunneling limit and 
the deviation of the splitting from unity is proportional to the 
square-root of the deviation of the conductance 
from unity in the strong tunneling limit.  This agrees
with the experimental data on double dot systems. 
We have also extracted the Luttinger parameters from the experiment
and have predicted  temperature power laws for the same set-up
which can be experimentally tested. 


\vskip -0.6 true cm


\begin{thebibliography}{99}

\vskip -1.4 true cm

\bibitem{GRABERT}
{\it Single charge tunneling}, edited by H. Grabert and M. H. Devoret, 
( Plenum Press, New York, 1992)

\bibitem{DD1}
F. R. Waugh {\it et al}, \prl{\bf 75}, 705 (1995); F. R. Waugh {\it et al},
\prb{\bf 53}, 1413 (1996).

\bibitem{DD2}
Molenkamp {\it et al}, \prl{\bf 75}, 4282 (1995); van der Vaart {\it et al}
\prl {\bf 74}, 4702;
R. H. Blick {\it et al}, \prb{\bf 53}, 7899 ((1996);
D. C. Dixon 
{\it et al}, \prb{\bf 53}, 12625 (1996); C. Livermore 
{\it et al},  Science {\bf 274}, 1332 (1996); Fujisawa {\it et al}, Science 
{\bf 282}, 932 (1998).

\bibitem{DD3}
For a review, see W. G. van der Wiel {\it et al}, cond-mat/0205350. 

\bibitem{CD}
I. M. Ruzin {\it et al}, \prb{\bf 45}, 13469 (1992); L. I. Glazman and
V. Chandrasekhar, Europhys. Lett. {\bf 19}, 623 (1992); C. A. Stafford
and S. Das Sarma, \prl{\bf 72}, 3590 (1994); G. Klimeck, G. Chen and
S. Datta, \prb {\bf 50}, 2316 (1994).

\bibitem{MGB} 
K. A. Matveev, L. I. Glazman and H. U. Baranger, \prb{\bf 53}, 1034 (1996);
{\it ibid} {\bf 54} 5637 (1996).

\bibitem{GH} J. M. Golden and B. I. Halperin, \prb{\bf 53}, 3893 (1996); 
{\it ibid} {\bf 65}, 115326 (2002).

\bibitem{LJ} S. Lamba and S. K. Joshi, \prb{\bf 62}, 1580 (2000).

\bibitem{MATVEEV}
K.A.Matveev, \prb{\bf 51}, 1743 (1995).

\bibitem{ASSA}
P. Rojt, Y. Meir and A. Auerbach, \prl{\bf 89}, 256401 (2002).


\bibitem{KF} C. L. Kane and M. P. A. Fisher, Phys. Rev. B {\bf 46}, 15233
(1992).

\bibitem{LAL} S. Lal, S. Rao and D. Sen, Phys. Rev. Lett. {\bf 87}, 026801 
(2001); {\it ibid}  Phys. Rev. B {\bf 65}, 195304 (2002).

\bibitem{RS} For a review, see S. Rao and D. Sen, cond-mat/005492, published
in `Field theories in  condensed matter physics, Ed. S. Rao, 
(IOP publications, U.K., May 2002).

\bibitem{FLENSBERG} K. Flensberg, Phys. Rev. B {\bf 48}, 11156 (1993).

\end{thebibliography}
\end{document}